\newcommand{\p}{\partial}

\newcommand{\n}{\nabla}
\newcommand{\ap}{\alpha'}
\newcommand{\bb}{\beta}
\newcommand{\RR}{I$\!\!\:$R}
\newcommand{\sra}{\stackrel{\rightarrow}}
\newcommand{\sla}{\stackrel{\leftarrow}}

\documentstyle[12pt]{article}

\begin{document}
\begin{flushright}
ITP-SB-93-08 \\
February 3, 1993
\end{flushright}
\begin{center}

\vskip3em
{\bf \huge
Wave-type Solutions in the Nonlinear $\sigma$-model with the Dilaton}

\vskip3em
{\large Eugene Tyurin}

\vskip1em

{Institute for Theoretical Physics, SUNY at Stony Brook \\
E-mail: gene@insti.physics.sunysb.edu}

\end{center}

\vskip 2cm

\begin{abstract}
We study a class of classical dilaton vacua in string theory that
depend on the light-cone variable $z=t\pm x$ and, thus, have wavelike
behavior. One of the interesting results is the existence of a solution
subclass with perfectly regular space-time geometry, where the string
coupling constant can be made arbitrarily small.
\end{abstract}
\pagebreak

\vspace{ 2 \baselineskip}
Classical solutions in the string theory correspond to conformal field
theories --- they are represented by the set of background fields that allows
the existence of the tree-level world-sheet (sphere in some chosen
parameterization). Thus these classical solutions can be called stringy vacua,
as loop corrections are identified with the quantum fluctuations of the
world-sheet.

In this paper, starting from one particular form of the space-time geometry,
we will find the wide class of the wave-type classical solutions that can
be made to have singular as well as regular behavior.

\vspace{\baselineskip}
We establish our notation. The $\sigma$-model action with the dilaton
term is
\begin{equation} \label{ssigma}
S_\sigma= \frac{1}{4\pi\ap} \int d^2 \sigma \sqrt{h} \left[ h^{ab} \p_a
X^m \p_b X^n g_{mn}(X^r) - \frac{\ap}{2} \phi(X^r) R^{(2)} \right],
\end{equation}
where $h^{ab}$ is the world-sheet metric, $g_{mn}$ is the metric on the
D-dimensional space-time with the coordinates $X^r$, $\phi (X^r)$ denotes
the dilaton field, and $R^{(2)}$ is the world-sheet curvature scalar.

To lowest order in $\ap$, the $\bb$-functions that should be equal to zero
to insure both conformal invariance and ultraviolet finiteness of the
action (\ref{ssigma}), are
\begin{eqnarray}
\bb^G_{mn} = & R_{mn} - \n_m \n_n \phi = 0 \label{bg} \\
\bb^{\Phi} = & R + (\n \phi)^2 - 2 \n^2 \phi + \Lambda = 0, \label{bf}
\end{eqnarray}
where $\Lambda=(D-D_{crit})/3\ap$ plays the role of ``cosmological constant"
in the low-energy effective action
\begin{equation}
S_{eff} = \int d^D X \sqrt{ g} e^{-\phi} \left[ R^{(D)} - (\n\phi)^2 +
\Lambda \right] \label{seff}
\end{equation}
Note that in general, the dilaton allows us to formulate the theory with the
target space dimensionality
{$D\not=D_{crit}$}, because without it, the field equations (\ref{bg}) and
(\ref{bf}) turn into
$$
\left\{ \begin{array}{c}
R_{mn} = 0 \\
R + \Lambda = 0
\end{array} \right.
\Longrightarrow \Lambda=0 \Leftrightarrow D = D_{crit}
$$

In this paper we do not discuss other possible background fields: axion and
tachyon, because (1) the axion can be easily generated by duality
transformations \cite{oddz}; (2) the tachyon is absent in superstrings.

\vspace{\baselineskip}
Recently a lot of attention has been given to the study of conformal field
theory  with the geometry of the target space
\begin{equation} \label{2d}
ds^2 = \mp dt^2 + a^2(t) d\theta^2 + \sum_{i=1}^{D-2}d\xi_i^2,
\end{equation}
and the dilaton $\phi (t)$, where $t$ is a Minkowskian or Euclidean time.
In this case $\bb$-functions (\ref{bg}), (\ref{bf}) have the following form:
\begin{equation} \label{1}
-\bb^G_{tt} = {{\ddot a}\over{a}} + \ddot \phi = 0
\end{equation}
\begin{equation} \label{2}
\bb^G_{\theta\theta} = {{\ddot a}\over{a}}+{{\dot a}\over{a}}{\dot \phi} = 0
\end{equation}
\begin{equation} \label{3}
\bb^\Phi = 2 {{\ddot a}\over{a}} - ({\dot \phi})^2 + 2 {\ddot
\phi} + 2  {{\dot a}\over{a}} {\dot \phi} + \Lambda = 0,
\end{equation}
where as usual ``dot" $\equiv \p/\p t$.
It turns out that there are only four nontrivial solutions for the
equations (\ref{1})--(\ref{3}):
\begin{eqnarray}
a(t) & = & \tan(\omega t), \cot(\omega t), \tanh(\omega t),
\coth(\omega t) \nonumber \\
\label{a(t)} \phi(t) & = & \phi_0 + \ln \bar{a}^2(t), \mbox{ where } \\
\bar{a}(t)& = & \sin(\omega t), \cos(\omega t), \sinh(\omega t),
\cosh(\omega t) \mbox{ correspondingly} \nonumber
\end{eqnarray}
As one can notice these solutions describe the black-hole type singularity
(cf. \cite{holes,radii})
and the choice of D determines the value of $\omega$ ($a(t)=1/t,
\phi(t)=\phi_0 + \ln t^2$ is possible only in the case of $\Lambda=0$
\cite{radii}).

An obvious generalization of the metric (\ref{2d})
(see, for example, paper \cite{tor} and references therein) is
\begin{equation} \label{2d2d}
ds^2 = - dt^2 + dx^2 + a^2(t) d\theta_1^2 + b^2(x) d\theta_2^2
+ \sum_{i=1}^{D-4}d\xi_i^2
\end{equation}
But it is still essentially two-dimensional. Another variant: curved inclusion
of the compactified coordinates discussed in the work \cite{radii}
\begin{eqnarray} \label{radii}
ds^2 = -dt^2 + \sum_{i=1}^{D-1} \left[ 2\pi r_i(t) \right]^2
dX_i^2, \\
X^i \equiv X^i + 1 \nonumber
\end{eqnarray}
doesn't bring us any drastically distinct solutions for the ``radii" $r_i(t)$.

\vspace{\baselineskip}
For these reasons we consider the following geometry of the space-time:
\begin{equation} \label{mymet}
ds^2 = - dt^2 + dx^2 + a^2(t,x) d\theta_1^2 + b^2(t,x) d\theta_2^2
+ \sum_{i=1}^{D-4}d\xi_i^2
\end{equation}
$$
\phi = \phi(t,x)
$$
Here we keep the main features of the equations
(\ref{2d}), (\ref{2d2d}) and (\ref{radii}) while looking for
the nontrivial both in $t$ and $x$ functional dependence of the metric
coefficients and the dilaton field.

Equations (\ref{bg}) have now the following form: 
\begin{equation} \label{btt}
-\bb^G_{tt} = {{\ddot a}\over{a}} + {{\ddot b}\over{b}} + \ddot\phi
= 0,
\end{equation}
\begin{equation}
-\bb^G_{xx} = {{a''}\over{a}} + {{b''}\over{b}} + \phi'' = 0,
\end{equation}
\begin{equation} \label{btx}
-\bb^G_{tx} = {{\dot a'}\over{a}} + {{\dot b'}\over{b}} + \dot\phi'
= 0,
\end{equation}
\begin{equation} \label{b11}
\bb^G_{\theta_1\theta_1} = {{\ddot a}\over{a}} - {{a''}\over{a}} +
{{a'b'}\over{ab}} - {{\dot a \dot b}\over{ab}} + {{\dot
a}\over{a}}\dot \phi - {{a'}\over{a}}\phi' = 0,
\end{equation}
\begin{equation} \label{b22}
\bb^G_{\theta_2\theta_2} = {{\ddot b}\over{b}} - {{b''}\over{b}} +
{{a'b'}\over{ab}} - {{\dot a \dot b}\over{ab}} + {{\dot
b}\over{b}}\dot \phi - {{b'}\over{b}}\phi' = 0,
\end{equation}
where ``prime" $\equiv\p/\p x$.
One can notice that the above system includes the two types of
partial differential equations and allows the following compact
representation:
\begin{equation} \label{this}
\left( {{\p^2 \phi}\over{\p X^i \p X^j}} + \sra{\p_i}\sra{\p_j} \right)
C^\kappa(t,x)C^\lambda(t,x) = 0 \qquad \kappa \not= \lambda,
\end{equation}
\begin{equation} \label{zero}
C^\kappa(t,x) \eta^{ij} \left[
\sra{\p_i} - \sla{\p_i} + {{\p \phi}\over{\p X^i}} \right] \sra{\p_j}
C^\lambda(t,x) = 0 \qquad \kappa \not= \lambda,
\end{equation}
where $C^{\kappa, \lambda}(t,x)$ denotes $a(t,x)$ or $b(t,x)$,
$X^i=\{t,x\}$ and $\eta^{ij} = diag\{-1, 1\}$. Unfortunately we do
not know the general solution for this problem and, though the proposed
geometry is an obvious generalization of (\ref{2d}), (\ref{2d2d}),
(\ref{radii}), such ``obvious" assumptions as
$a,b = f(t) g(x) \mbox{ or } \left[ f(t) + g(x) \right]^p$
do not work unless they accidentally fall in the solution class that will be
introduced below.

\vspace{ \baselineskip}
Suppose $a, b, \phi$ to depend only on the single light-cone variable
{$z=t+x$} or $\tilde z=t-x$. All the subsequent formulae do not depend on the
choice of $z$ or $\tilde z$, so for the sake of clarity we will speak in the
terms of $z$.
With the above ansatz equations (\ref{zero}) are satisfied
identically for any $a, b, \phi$ and the three equations denoted as
(\ref{this}) are equivalent to each other, and are just
$$
{{a''(z)}\over{a(z)}} + {{b''(z)}\over{b(z)}} + \phi''(z) = 0,
$$
where ``prime" from now on means $\p/\p z$. The only component of the Ricci
tensor that is not equal to zero identically is \begin{equation} \label{Rzz}
R_{zz} = - {{a''}\over{a}} - {{b''}\over{b}},
\end{equation}
hence $R={R_a}^a\equiv 0$. One can also check that both $\n^2 \phi \equiv 0$
and
$(\n \phi)^2 \equiv 0$. Because of these cancellations, equation (\ref{bf})
is equivalent to the condition $\Lambda=0 \Leftrightarrow D=D_{crit}$
and this is distinct from the previously discussed cases.

So the full system of the lowest order $\bb$-function equations we should
satisfy is
\begin{equation}
 {{a''}\over{a}} + {{b''}\over{b}} + \phi'' = 0, \label{ooo}
\end{equation}
\begin{equation}
 \Lambda = 0
\end{equation}

Since the equations (\ref{this}) and (\ref{1}) are similar to each
other, one can employ as the wave-type solutions all of the options
(\ref{a(t)}) which should be identified now with moving singularities.

Generally, we are not interested in the higher orders in $\ap$,
because in the supersymmetric theory one can always make the quasi-classical
approximation to be exact, but it is nice to notice that the second-order
corrections to the $\bb$-functions that are proportional to
$\delta \bb_{mn} =\ap R_{mabc} {R_{n}}^{abc}$ are identically equal to zero
in our case.

Another interesting feature of the solution class we have found is that we can
always change the constant and linear powers of the dilaton field (there are no
first powers of the ordinary derivatives involved).

Of course, since we have only one equation for three functions, there is a
tremendous freedom of choice. For example, after fixing the dilaton field,
we obtain the following equations for $a(z),b(z)$ with an absolutely arbitrary
function $f(z)$:
\begin{eqnarray} \label{f(z)}
{{a''}\over{a}} + \phi'' - f = 0, \\
{{b''}\over{b}} + f = 0, \nonumber
\end{eqnarray}
that have the general form of $y''(x) + w(x)y(x) = 0$.
This type of differential equations has been thoroughly studied in
mathematics (for a review see, for example, \cite{kamke})
and here we will just say that the existence of a finite or infinite number
of zeros for $y(x)$ (i.e. metric irregularities in our terminology) is
guaranteed for a wide class of $w(x)$.

Since the ultimate purpose of the model-building approach that we employ here
is to obtain something that seems to be a reasonable choice of the classical
vacuum in the string theory, it seems to be interesting to find
solutions with a ``perfect behavior". Recall that, since we work in the
perturbation theory with the effective string coupling constant g
$=exp(\phi/2)$,
and the dilaton $\phi(z)$ can always be shifted by some finite constant
$\phi_0$, g can always be made arbitrarily (but, in general, not infinitely)
small if $\phi(z)\not= +\infty$ for all $z$. Additionally we will require the
target space metric $g_{mn}$ to be regular everywhere except perhaps
light-cone infinity. The following statement can point to some of these
``nice" solutions.

\noindent
\underline{Theorem} Let $a(z) \in C^2($\RR$)$ be a symmetric, monotonically
non-increasing for $z>0$ function and let
$$ \lim_{z \to 0} a(z) = c_1 < \infty \qquad
\lim_{z \to +\infty} a(z)=c_2 \geq 0, \mbox{ then the dilaton } $$
$$\phi_a(z)= - \int \int {{a''}\over{a}} dz dz \to
-\infty \mbox{ as } z \to +\infty $$
and $\phi_a(z) < +\infty$ for all z. \\
\underline{Proof}
First of all, note that from the condition $a(z)=a(-z)$ and from the
continuity of $a'(z)$ follows that there exists a point $\zeta>0$ such that
$$ a''(z) \left\{ \begin{array}{c}
< 0 \mbox{  } z < \zeta \\
> 0 \mbox{  } z > \zeta
\end{array} \right., \quad
\mbox{then } \phi_a''=-{{a''}\over{a}} \left\{
\begin{array}{c}
> 0 \mbox{  } z < \zeta \\
< 0 \mbox{  } z > \zeta
\end{array} \right.
$$
It is clear that $\phi_a''$ cannot be positive for $z>\zeta$, because
$\lim_{z \to +\infty} a''(z) =-0$. On the next step, there is such
$\gamma>\zeta$, that
$$ \phi_a'=\int dz \phi_a'' \left\{ \begin{array}{c}
> 0 \mbox{  } z < \gamma \\
< 0 \mbox{  } z > \gamma
\end{array} \right. $$
Again, $\phi_a'$ cannot be positive for $z>\gamma$, because of the infiniteness
of the interval where $\phi_a''<0$. Finally, we find that
$\lim_{z \to +\infty} \phi_a(z) =-\infty$, because $\phi_a'<0$ for $z>\gamma$.
{\bf Q.E.D.}

Note that if we consider monotonically increasing $a(z)$ for $z>0$ then
$\lim_{z \to +\infty} \phi_a(z) =+\infty$. Unfortunately it is impossible to
analyse oscillating, non-monotonic functions in the same fashion. But,
with the theorem proven above, one can pick a lot of metrics that
contain this ``nice" dilaton background (cf. (\ref{f(z)}): of course, we assume
here that $f(z)=-b''/b= \phi_b''$).

\vspace{\baselineskip}
We have discussed the class of the classical dilaton solutions to the
critical string theory that allows a wave-packet interpretation.
These solutions allow both singularities and finite, smooth backgrounds.
Since for the
space-time geometry we have considered, there are some miraculous cancellations
in the $\bb$-function equations, it may be interesting to investigate this
topic for some more general initial assumptions.
The vacua presented in this paper can also be found in
\cite{prop}, where the metrics of the form (\ref{radii}), but in the light-cone
variables were studied.

\vspace{ 2 \baselineskip}
I would like to express my deep gratitude to M. Ro\v{c}ek for suggesting
this problem to me and for his advice.

\vspace{ 2 \baselineskip}

\end{document}